\documentclass[aps,nofootinbib,prd,preprintnumbers,showpacs]{revtex4}
\usepackage{wrapfig}
\usepackage{amsfonts}
\usepackage{amsmath}
\usepackage{hyperref}
\usepackage{amssymb}
\usepackage[english]{babel}
\usepackage{graphicx}
\usepackage{epsfig}
\usepackage{bm}
\usepackage{longtable}
\usepackage{verbatim}
\usepackage{longtable}
\usepackage{color}
\usepackage{subfigure}
\usepackage[utf8]{inputenc}
\usepackage{xcolor}
\usepackage{xspace}
\usepackage{ulem}
\usepackage{lscape}
\usepackage{wasysym}
\usepackage[toc,page]{appendix}
\usepackage{mathtools}
\usepackage{multirow,rotating}
\usepackage{stackrel}
\usepackage{cancel}
\usepackage{pifont}
\usepackage{dcolumn}

\newcommand{\mathsym}[1]{{}}

\begin{document}

\title{
Background for gravitational wave signal at LISA from refractive index of solar wind plasma
}

\author{Adam Smetana,\thanks{E-mail: adam.smetana@cvut.cz}}
\affiliation{Institute of Experimental and Applied Physics, Czech Technical University in
Prague, Prague, Czech Republic}



\begin{abstract}
\noindent
\textbf{Abstract:} A strong indication is presented that the space-based gravitational antennas, in particular the LISA concept introduced in 2017 in response to the ESA call for L3 mission concepts, are going to be sensitive to a strong background signal interfering with the prospected signal of gravitational waves. The false signal is due to variations in the electron number density of the solar wind, causing variations in the refractive index of plasma flowing through interplanetary space. As countermeasures, two solutions are proposed. The first solution is to deploy enough solar wind detectors to the LISA mission to allow for reliable knowledge of the solar wind background. The second solution is to equip the LISA interferometer with a second laser beam with a distinct wavelength to allow cancelling of the background solar wind signal from the interferometric data.
\end{abstract}


\maketitle



\section{Introduction}

Interferometers are highly appreciated in physics for their excellent sensitivity to the signal of interest. Examples of their common and old usages are the Rayleigh interferometer for measurement of refractive index of a transparent gas, or the Michelson interferometer designed to detect the Earth's motion through the supposed luminiferous aether. At the cutting edge of Michelson laser interferometry are the ground-based gravitational wave detectors LIGO \cite{Abramovici1992325} and VIRGO \cite{Acernese20021421}, which reached unprecedented sensitivity for displacement of their four-kilometers-long arm ends at the current upgraded level of $\sim10^{-18}\mathrm{m}$ in the gravitational wave frequency range $\sim10-1000\,\mathrm{Hz}$. That allowed for the first-ever Nobel-Prize observation of gravitational waves from collapsing mergers of celestial origin in 2015 \cite{LIGO_Virgo}. The follow-up experiment, the Laser Interferometer Space Antenna (LISA) with its latest concept introduced in 2017 in response to the ESA call for L3 mission concepts \cite{LISA}, is planned to be space-based laser interferometer with significantly larger arm length of $2.5\times10^6\,\mathrm{km}$ allowing for sensitivity at the level of $\sim10^{-12}\mathrm{m}$ in the low frequency range $10^{-4}-10^{-1}\,\mathrm{Hz}$, opening the window for the heaviest and most diverse gravitational-wave sources which are inaccessible from the ground. 

Placing the gravitational wave observatory into space is motivated by dropping the necessity to construct high volume vacuum tubes, by lack of limitations on the length of the arms of the interferometer, and by avoiding the jitter noise of seismic origin. On the other hand, the ground-based facilities allow for controlling the vacuum impurity and its time stability. Mainly the latter is important to avoid spurious signals as pointed out in this work. In this work the effect of vacuum impurity of the interplanetary space environment of the LISA caused by time-variable solar activity has been studied. Namely, under the assumption of validity of the Sellmeier equation for the refractive index of plasma, extrapolated to extremely small electron densities, it is shown that the solar wind plasma density at the $1\,\mathrm{AU}$ solar orbit provides small deviation of electromagnetic wave phase velocity from its pure vacuum value. The time variations of the solar wind electron density produces an accountable signal at gravitational-wave antennas of the sensitivity proposed for LISA. The resulting large background for significant portion of the prospected gravitational wave signals of interest would jeopardize the scientific mission of the LISA. Possible countermeasures are proposed.

\section{Refractive index of solar wind plasma}

The solar wind composition is dominated by electrons and protons. Smaller component of solar wind is given by $\alpha$-particles and heavier ions. The quasi-neutrality of the solar wind is guaranteed by comparable number densities of electrons and protons. The solar wind at the distance of $1\,\mathrm{AU}$ from the Sun has the average particle density $\sim10\,\mathrm{cm}^{-3}$. At the scales of the optical wavelength range, the solar wind is practically a collision-less ideal plasma. Therefore, for the purpose of this work, its electron component can be treated as a gas of free electrons characterized by the plasma frequency 
\begin{equation}\label{omegap}
    \omega_{p}^2=\frac{n_e e^2}{\epsilon_0 m_e}\,,
\end{equation}
where $n_e$ is the electron number density, $m_e$ is the electron mass, $e$ is the value of the elementary electric charge and $\epsilon_0$ is the permittivity of vacuum. This situation is analogous to the ionospheric plasma, where the index of refraction is given by the well-known Sellmeier equation \cite{Darwin}. Due to the inverse-power dependence on the mass of the plasma particles, contributions from protons and ions can be neglected. Extrapolating its validity to the range of near-optical wavelengths and into the extremely low electron densities of solar wind, the formula for refractive index $n$ of the solar wind plasma can be used
\begin{equation}\label{n}
    n=\sqrt{1-\frac{\omega_{p}^2}{\omega^2}}\doteq 1-\frac{1}{2}\frac{\omega_{p}^2}{\omega^2}+\dots\,,
\end{equation}
where $\omega$ is the angular frequency of the electromagnetic wave passing through the plasma. The assumption of validity of the formula Eq.~\ref{n} is justified by negligible values of gyrofrequencies of the solar wind plasma in comparison with $\omega_p$ and $\omega_p\ll\omega$. For the values of the electron density $\sim10\,\mathrm{cm}^{-3}$ and for $\omega$ in the optical range, the deviation of $n$ from unity is truly negligible, at the level of $\sim10^{-21}-10^{-20}$, which would require measurement by an instrument of comparable sensitivity.

\section{LISA parameters}

The space-based gravitational wave antenna LISA is proposed to be based on a system of free-falling test masses in solar orbit, which serve as both geodetic reference and mirrors for laser beams at the ends of the interferometer arms. The test masses are shielded against the space weather and against the solar wind by bodies of drag-free spacecraft, which absorb a major portion of momentum transfer from the environment by constantly adjusting their trajectories to the trajectories of the test masses. The solar activity and solar wind as possible sources of noise for the LISA has been analysed in \cite{solar_LISA}. Due to enormous effort in suppressing LISA's noise level \cite{LISA}, the displacement noise linear spectral density, $\sqrt{S_\mathrm{IFO}}$, of the interferometric test-mass-to-test-mass distance has been proposed to be 
\begin{eqnarray}\label{SIFO}
    &&\sqrt{S_{\mathrm{IFO}}}\leq 10^{-11}\frac{\mathrm{m}}{\sqrt{\mathrm{Hz}}}\sqrt{1+\left(\frac{2\mathrm{mHz}}{f}\right)^4} \nonumber \\
    &&\mathrm{for}\ \ \ 10^{-4}\mathrm{Hz}\leq f\leq 10^{-1}\mathrm{Hz} \,,
\end{eqnarray}
where $f$ is the frequency parameter spanning the gravitational wave frequency range of interest. Such a noise level has been required in order to be smaller than the strain level of the signal of interest from the gravitational waves and not to cover it. That was successfully demonstrated and even improved by the the LISA Pathfinder single-test-mass mission \cite{LISAPathfinder}. 

The proposed interferometric configuration of the LISA observatory is based on three arms with six active laser links operating on the wavelength
\begin{equation}\label{laser}
    \lambda = 1064\,\mathrm{nm}\,,
\end{equation} 
between three identical spacecraft in an equilateral triangular formation, each separated by a mean distance of
\begin{equation}
    L = 2.5\times10^6\,\mathrm{km}\,,
\end{equation} 
and carrying two identical test masses each. The LISA reference orbit is a stable Earth-trailing heliocentric orbit about 50 million km from Earth.

The amplitudes of typical gravitational wave signals for the LISA with the proposed parameters can be found in the LISA proposal \cite{LISA}, or in more detail in \cite{Moore_2014}.


\section{Solar wind data}

For the present study, the relevant solar wind data are that of the electron number density $n_e$ with sufficient sampling frequency $\gtrsim0.1\,\mathrm{Hz}$. This requirement has been met by the WIND mission. The WIND is a spin stabilized spacecraft, which has been several times placed into the halo orbit around the Earth's L1 Lagrange point to observe the unperturbed solar wind that is about to impact the magnetosphere of Earth. Data from the WIND's halo orbit period from February 1997 till November 1998 (https://pwg.gsfc.nasa.gov/istp/wind/Timeline.htm) are used. All the data used in the analysis comes from the WIND/SWE electron instrument \cite{wind}. The WIND/SWE team claims that the electron instrument response has varied over time. There are periods when the derived electron density may be as much as a factor 2 too low. Although the electron density is not derived absolutely, relative changes in electron density can usually be relied on, which is the key feature for this analysis. 

\begin{figure}[h]
	\includegraphics[width=1.0\linewidth]{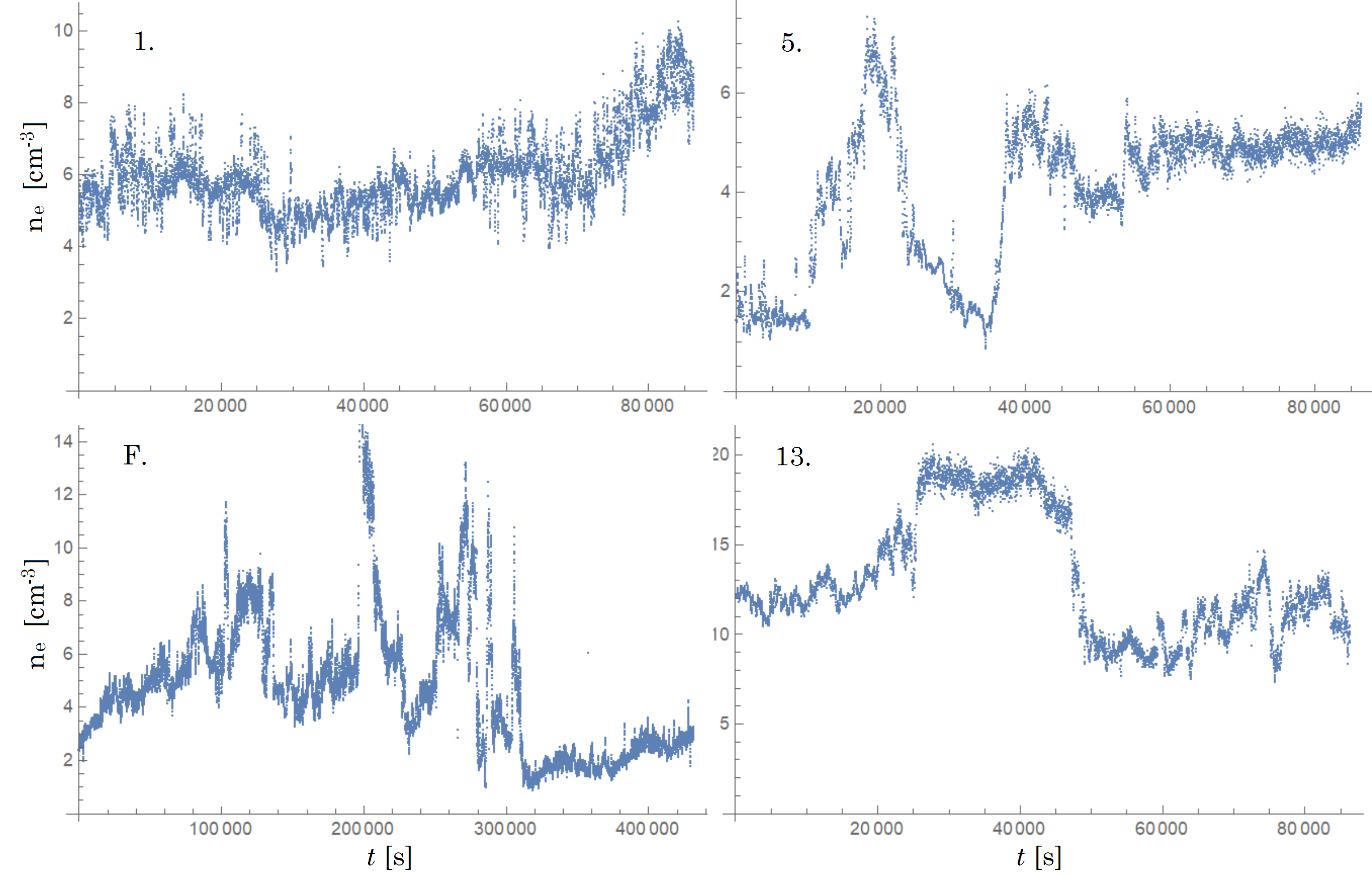}
    \caption{The upper panels show two typical calm-condition ($\overline{n_e}<8.0$) 24-hours data samples of the solar wind electron density. The lower panels show two typical data samples of the solar wind electron density during the solar storm events or active conditions ($\overline{n_e}>10.0$).}
    \label{WIND_spectra}
\end{figure}

On the Fig.~\ref{WIND_spectra} the examples of data used for the analysis are displayed. They are selected quite arbitrarily, just avoiding some obvious discontinuities and outliers, and keeping only data collected far from the Earth's magnetosphere. As the frequencies relevant for LISA range down to $0.1\,\mathrm{mHz}$ at least $1$ day long data samples are taken into the analysis to cover at least $\sim10$ periods. The selection has been done to include some of the calm space conditions data as well as some of the solar storm events data. In the Tab.~\ref{WIND_spectra_list_random} and Tab.~\ref{WIND_spectra_list_events} the data samples with their main characteristics are listed. It shows the mean, $\overline{n_e}$, maximum, $n_{e}^\mathrm{max}$, and minimum, $n_{e}^\mathrm{min}$, values of each data set. The data samples used in Tab.~\ref{WIND_spectra_list_random} are of 24h duration, chosen arbitrarily to be from the 11th day of each month. The data from the Tab.~\ref{WIND_spectra_list_events} cover periods pre-selected by the WIND collaboration to cover certain significant solar events.

\begin{table}
	\caption{List of data samples arbitrarily chosen from the period Feb 1997 - Nov 1998. Initial time is always 00:00:00.}
	\label{WIND_spectra_list_random}
	\begin{tabular}{rlccccl} 
		\hline
		no. & initial date & length & $\overline{n_e}$ & $n_{e}^\mathrm{max}$ & $n_{e}^\mathrm{min}$ & $[\mathrm{cm}^{-3}]$ \\
		\hline
		1. & 1997/07/11 & 24\,h & $5.9$ & $10.0$ & $3.0$& \\
		2. & 1997/08/11 & 24\,h & $6.4$ & $10.5$ & $3.5$& \\
		3. & 1997/09/11 & 24\,h & $4.4$ & $7.3$ & $2.7$& \\
		4. & 1997/10/11 & 24\,h & $7.6$ & $13.5$ & $4.0$ &\\
		5. & 1997/12/11 & 24\,h & $4.0$ & $7.2$ & $0.8$ &\\
		6. & 1998/01/11 & 24\,h & $7.4$ & $10.3$ & $4.6$ &\\
		8. & 1998/03/11 & 24\,h & $3.4$ & $4.5$ & $2.6$ &\\
		9. & 1998/04/11 & 24\,h & $5.6$ & $9.2$ & $2.2$ &\\
		10. & 1998/05/11 & 24\,h & $3.3$ & $4.4$ & $2.3$ &\\
		11. & 1998/06/11 & 24\,h & $6.6$ & $13.5$ & $0.1$ &\\
		12. & 1998/07/11 & 24\,h & $5.3$ & $11.4$ & $1.4$ &\\
		13. & 1998/09/11 & 24\,h & $13.1$ & $20.6$ & $7.4$ &\\
		14. & 1998/10/11 & 24\,h & $4.0$ & $6.0$ & $2.4$ &\\
		\hline
	\end{tabular}
\end{table}

\begin{table*}
	\caption{List of solar events data samples chosen from the period Feb 1997 - Nov 1998. Initial time is always 00:00:00.}
	\label{WIND_spectra_list_events}
	\begin{tabular}{rlcccccl} 
		\hline
		no. & event & initial date & length & $\overline{n_e}$ & $n_{e}^\mathrm{max}$ & $n_{e}^\mathrm{min}$ & $[\mathrm{cm}^{-3}]$ \\
		\hline
		A. & ISTP-SEC & 1997/11/04 & 24\,h & $6.1$ & $12.0$ & $4.0$ &  \\
		B. & ISTP-SEC CME & 1998/01/21 & 24\,h & $8.0$ & $22.0$ & $3.5$ &  \\
		C. & ISTP-SEC CME & 1998/04/15 & 24\,h & $4.2$ & $7.2$ & $3.1$ &  \\
		D. & ISTP-SEC CME-Geostorm & 1998/04/29 & 96\,h & $5.4$ & $15.0$ & $1.2$ &  \\
		E. & ISTP-SEC Solar-GeoStorm & 1998/09/23 & 96\,h & $5.5$ & $15.0$ & $0.5$ &  \\
		F. & ISTP-SEC Solar-GeoStorm & 1998/08/24 & 120\,h & $4.7$ & $14.5$ & $1.0$ &  \\
		\hline
	\end{tabular}
\end{table*}

\section{Estimation of the background for LISA from the solar wind plasma effect }

Based on the Eq.~\ref{n} and Eq.~\ref{omegap}, the deviation of refractive index depends linearly on the electron number density, which is the only source of its time variation. The time variation of the refractive index can be described in terms of time varying effective displacement 
\begin{equation}\label{hSW}
    h_{\mathrm{SW}}(t)=\frac{1}{2}\frac{n_e(t)e^2}{\epsilon_0 m_e\omega^2}L\,,
\end{equation}
mimicking the signal of displacement due to the gravitational wave distortion of the space fabric. 

\begin{figure*}
    \centering
	\includegraphics[width=1.0\linewidth]{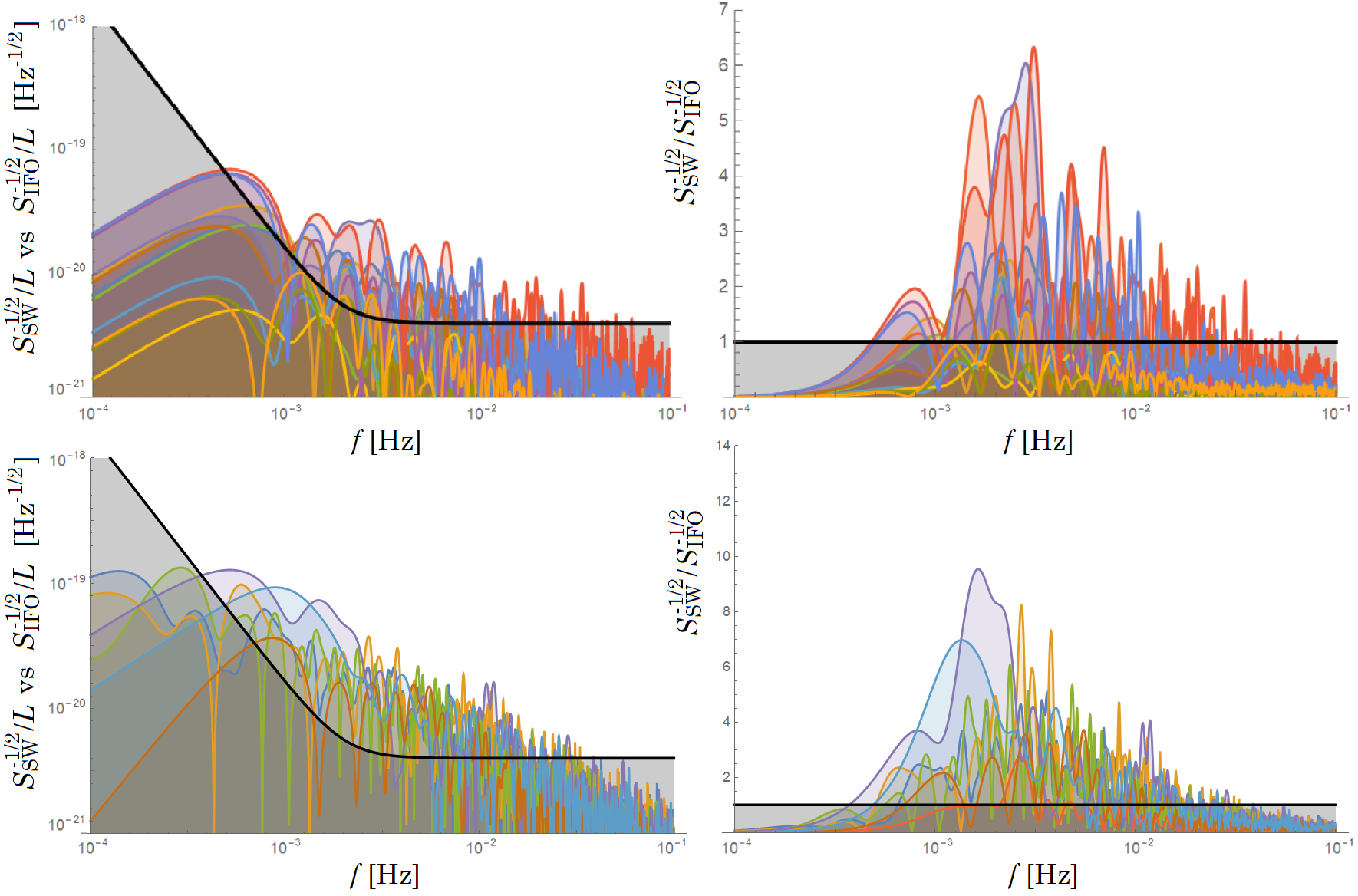}
   \caption{In black (thick) the proposed required sensitivity performance of the LISA observatory, $\sqrt{S_{\mathrm{IFO}}}$, given in Eq.~\ref{SIFO}, is shown. The noise level should not exceed the black noise level. In color (thin) the linear spectral densities of the effective displacement, $h_{\mathrm{SW}}(t)$ introduced in the Eq.~\ref{hSW}, are shown, representing the solar wind plasma signal due to the variations of the solar wind electron number densities. In the left column the comparison of the displacement linear spectral density relative to the arm length $L$ is shown. In the right column the ratio of the solar wind effective displacement linear spectral density and the required sensitivity performance is shown. In the upper row the signals of the calm-condition data samples (1.--12.,14.) from Tab.~\ref{WIND_spectra_list_random} are shown, and in the lower row the signals of the solar storm events (A.--F.) from the Tab.~\ref{WIND_spectra_list_events} and active-condition data samples (13.) from  Tab.~\ref{WIND_spectra_list_random} are shown. }
    \label{comparison}
\end{figure*}

In order to compare the solar wind signal with the required strain sensitivity of LISA, $\sqrt{S_{\mathrm{IFO}}}$ given in Eq.~\ref{SIFO}, the linear spectral density $\sqrt{S_{\mathrm{SW}}}(f)$ for the effective displacement $h_{\mathrm{SW}}(t)$ is calculated. The result is shown in the Fig.~\ref{comparison} plotting the linear spectral densities $\sqrt{S_{\mathrm{SW}}}$ for all selected data samples together (in colors) with the required sensitivity $\sqrt{S_{\mathrm{IFO}}}$ (in black). In the frequency range
\begin{eqnarray}
    10^{-3.3}\mathrm{Hz}\leq f\leq 10^{-1.4}\mathrm{Hz} &\mathrm{\ for\ }&\mathrm{calm\ conditions} \\
    10^{-3.5}\mathrm{Hz}\leq f\leq 10^{-1.0}\mathrm{Hz} &\mathrm{\ for\ }&\mathrm{active\ conditions} 
\end{eqnarray}
the solar wind signal in the analysis has amplitude greater than the required noise level, $\sqrt{S_{\mathrm{IFO}}}$, by factor up to $\sim6$ (calm conditions) around the frequency $10^{-2.6}\mathrm{Hz}$, or by factor up to $\sim10$ (active conditions) around the frequency $10^{-2.4}\mathrm{Hz}$. Out of all used data sets only a single one (no.~8.) provides signal, which does not exceed the required noise level, $\sqrt{S_{\mathrm{IFO}}}$, and three (no.~3.,10.,14.) only very little. 

The result of the analysis indicates that the solar wind causes a significant background for the expected gravitational wave signal $\sqrt{S_{\mathrm{h}}}$, noted in the LISA proposal \cite{LISA}. This result indicates a severe constraint of the scientific performance of the proposed LISA mission.

\section{Discussion of the result and Proposal of countermeasures to safe the LISA}

The danger of the signals from solar wind plasma effects observable by LISA are applicable to any space-based interferometric observatory sensitive to the sub-Hz frequency range used here. It shows the weak point of replacing the laboratory vacuum by the vacuum of the interplanetary open space, which exhibits uncontrollable time variations of the solar wind plasma density caused by solar activity. The characteristic frequencies of the solar activity span a large range covering the frequency range of LISA's interest. The present analysis, based on the formulas Eq.~\ref{omegap} and Eq.~\ref{n}, indicates that the solar wind plasma effect on the refractive index of the LISA's space environment for the wavelength of its laser is strong enough to interfere with expected gravitational wave signals.

The result presented in this work is just an estimate of the effect. The main uncertainty of the result lies in our ignorance about the space variation of the solar wind density. The available data carries the information only about the time variation being collected in a single spot - the WIND spacecraft position - at a given time. If the space variation of the solar wind is dominated by a characteristic wavelength $\lambda_\mathrm{SW}$
\begin{equation}
    \lambda_\mathrm{SW}\ll L\,,
\end{equation}
then the solar wind electron density variations would be averaged out along the laser trajectory of the interferometric arm, effectively suppressing the solar wind signal. 

Similar effective suppression of the solar wind signal is expected for the arm in the case it appears parallel with the solar wind velocity vector at some instances on the orbit. Taking the typical value of the solar wind velocity to be $v_\mathrm{SW}\sim500\,\mathrm{km/s}$, the integration over the time $t=L/v_\mathrm{SW}\sim5000\,\mathrm{s}$ of the solar wind electron density has to be performed, effectively suppressing the solar wind signal over almost entire frequency range of interest. However this condition is fulfilled only for limited periods and never for all three arms at once. 

At least at the current stage of development, it seems to be too risky to rely on the assumption that the averaging over space and time would save the LISA mission.

Clearly, the reliable measurement of the solar wind electron number density along the arms of the LISA interferometer would open the possibility to know and subtract the solar wind signal from the delivered data. Modelling of the LISA's solar wind plasma environment is subject to further detailed analysis, to see whether it would be sufficient to place solar wind detectors just on board of each of the three spacecraft, or it would be necessary to have more detailed data with shorter distance sampling along each of the arms. The latter seems more likely as suggested by significantly different solar wind data delivered by three spacecraft currently in orbit about the L1 point, see e.g. \cite{Walsh2019}. 

Another, much more precise option to subtract the solar wind signal is based on the fact that the deviation of the refractive index depends on the wavelength of the laser, see Eq.~\ref{hSW}. LISA could be equipped by second laser of distinct wavelength  
\begin{equation}
    \lambda'=r\lambda\,,
\end{equation}
using the same optical system including the same test-mass mirrors, where $\lambda$ is the wavelength of the originally proposed laser given in Eq.~\ref{laser} and the factor $r$ is their smartly chosen ratio different from unity. Such an improvement to LISA would allow the subtraction of the two data sets, $h(t)$ and $h'(t)$, corresponding to the laser wavelengths $\lambda$ and $\lambda'$, respectively, in the following way
\begin{equation}
    \Delta h(t) = r^2 h(t)-h'(t)\,.
\end{equation}
The resulting data set $\Delta h(t)$ is effectively free of the solar wind signal. Under the assumption that the gravitational wave signal scales with the laser wavelength only very weakly (or at least differently than the solar wind signal), the data set $\Delta h(t)$ still carries desired information. Of course, by such manipulation with data, the noise level and the systematic uncertainties increase.

\section{Conclusions}

This brief letter reports about the indication for significant sensitivity of space-based gravitational antennas, in particular of the LISA concept introduced in 2017, to the time-variable solar wind background, with signal comparable to that prospected for gravitational waves. Namely, it has been shown that the solar wind plasma density at the $1\,\mathrm{AU}$ solar orbit provides small time variations of the refractive index for the laser electromagnetic wave from its pure vacuum value, which would be effectively measured as a displacement of the test masses. It has been shown that the signal from the solar wind may provide a measurable background for a significant portion of the prospected gravitational wave signal of interest. This result severely constrains the scientific performance of the proposed LISA mission.

Two possible solutions have been proposed. The first solution is to deploy enough solar wind detectors to the LISA mission allowing for reliable knowledge of the background. The second solution is to equip the LISA interferometer with a second laser beam at a different wavelength to allow the cancellation of the solar wind signal from the combined measurement taken at each wavelength. The improved LISA mission would gain an added value of providing an excellent tool for research of solar wind plasma and would generate a valuable and unique data for the plasma theory in extreme conditions.

The purpose of this letter is to bring attention to this phenomenon and encourage the community to consider it when designing the LISA mission. More detailed analysis is needed, including the reliable 4D simulations of the space weather environment of the LISA and including theoretical studies and evaluations of the full variability of the solar wind plasma effects. It may turn out, after all, that the effect of the solar wind plasma will not be so harmful as indicated here, e.g., thanks to the averaging the solar wind plasma density variations out of the frequency range of the LISA's interest.

\section*{Acknowledgements}

I acknowledge the support of the Institute of Experimental and Applied physics, Czech Technical University in Prague, for many years of support. The work is dedicated to my Teacher.

\section*{Data availability}

The data underlying this article are available in Goddard Space Flight Center - Space Physics Data Facility (https://cdaweb.gsfc.nasa.gov/istp\_public/), at https://dx.doi.org/10.1007/BF00751326.




\bibliographystyle{mnras}
\bibliography{refLISA} 


\end{document}